\documentstyle[12pt]{article}
 at 14.4truept at 12.0truept
\title{ Deriving exact prepotential for $N = 2$ supersymmetric Yang-Mills 
theories from superconformal anomaly
with rank two  gauge groups}
\author{
{\large Satchidananda  Naik} \thanks{e-mail: naik@mri.ernet.in}
\\
  Mehta Research Institute of
 Mathematics \\
 and Mathematical Physics \\
 Chhatnag Road, Jhusi  \\
Allahabad-211 019, INDIA\\}

\begin{document}
\maketitle

\hspace*{\fill}

\hspace*{\fill}
\newcommand{\bee}{\begin{equation}}
\newcommand{\nn}{\nonumber}
\newcommand{\ee}{\end{equation}}
\newcommand{\ba}{\begin{array}}
\newcommand{\ea}{\end{array}}
\newcommand{\bea}{\begin{eqnarray}}
\newcommand{\eea}{\end{eqnarray}}
\newcommand{\ki}{\chi}
\newcommand{\eps}{\epsilon}
\newcommand{\pa}{\partial}
\newcommand{\lb}{\lbrack}
\newcommand{\Se}{S_{\rm eff}}
\newcommand{\rb}{\rbrack}
\newcommand{\de}{\delta}
\newcommand{\th}{\theta}
\newcommand{\rh}{\rho}
\newcommand{\ka}{\kappa}
\newcommand{\al}{\alpha}
\newcommand{\bt}{\beta}
\newcommand{\si}{\sigma}
\newcommand{\bsi}{\Sigma}
\newcommand{\vp}{\varphi}
\newcommand{\gm}{\gamma}
\newcommand{\gb}{\Gamma}
\newcommand{\om}{\omega}
\newcommand{\et}{\eta}
\newcommand{\gt}{ {g^2 T }\over{4 {\pi}^2}}
\newcommand{\qab}{{{\sum}_{a\neq b}}{{q_a q_b}\over{R_{ab}}}}
\newcommand{\omb}{\Omega}
\newcommand{\pr}{\prime}
\newcommand{\ra}{\rightarrow}
\newcommand{\nb}{\nabla}
\newcommand{\MSb}{{\overline {\rm MS}}}
\newcommand{\lnh}{\ln(h^2/\Lambda^2)}
\newcommand{\cz}{{\cal Z}}
\newcommand{\h}{{1\over2}}
\newcommand{\Lm}{\Lambda}
\newcommand{\inft}{\infty}
\newcommand{\abschnitt}[1]{\par \noindent {\large {\bf {#1}}} \par}
\newcommand{\subabschnitt}[1]{\par \noindent
                                          {\normalsize {\it {#1}}} \par}
\newcommand{\skipp}[1]{\mbox{\hspace{#1 ex}}}
 
%
%
%
%
\newcommand\dsl{\,\raise.15ex\hbox{/}\mkern-13.5mu D}
\newcommand\delsl{\raise.15ex\hbox{/}\kern-.57em\partial}
\newcommand\Ksl{\hbox{/\kern-.6000em\rm K}}
\newcommand\Asl{\hbox{/\kern-.6500em \rm A}}
\newcommand\Dsl{\hbox{/\kern-.6000em\rm D}} 
\newcommand\Qsl{\hbox{/\kern-.6000em\rm Q}}
\newcommand\gradsl{\hbox{/\kern-.6500em$\nabla$}}
\newpage
\begin{abstract} \normalsize
  The exact prepotential for $N = 2$ supersymmetric Yang-Mills theory is derived
from  the superconformal anomalous Ward identity for the gauge group $SU(2)$ and 
$SU(3)$
which can be generalized to  any  other rank two gauge group.
\end{abstract}

\vskip10.0cm

\newpage
\pagestyle{plain}
\setcounter{page}{1}
\setcounter{section}{1}
\abschnitt{1. Introduction}

There is  a spectacular progress in understanding the non-perturbative
behaviour of supersymmetric Yang-Mills theories in recent past \cite{SW}.
In their pioneering work    Seiberg and Witten  obtained an exact low energy 
Wilsonian effective action for $N = 2$ supersymmetric Yang-Mills theory for the
 $SU(2)$ gauge group. Due to $N = 2 $ SUSY the effective action has
 to be holomorphic
and its perturbative one loop $\bt $ function is exact.
 The quantum moduli space of 
this theory is described as a one parameter family of elliptic curves.
 The period 
of this curve fixes the metric on the quantum moduli
 space as well as the spectrum 
of BPS states.
Then making use of the electric-magnetic duality i.e. strong coupling 
behaviour is systematically related to the weak coupling behaviour of
 the dual fields,
the exact prepotential is obtained.
 This was immediately generalized to theory with 
arbitrary $SU(N)$   gauge group \cite{KLSY, AF} and with matter multiplet
 \cite{HN}.

Later on   Boneli  et. al.  and Matone   \cite{BM} have shown 
that the moduli space is equivalent
to genus zero Rieman surface and $ u = <tr {\phi}^2> $
( where $\phi$ is the scalar 
component of the $N = 2$ SUSY ) is the uniformizing cordinate of the moduli 
space. In this way without using the arguments of duality, 
they obtain the effective potential which is same as that of 
Seiberg and Witten.
Recently in two consecutive works Magro et al.\cite{Mag} and Flume et al.
\cite{Flu}  have obtained the 
same result without making use of duality. Their main argument is
that the moduli space is the space of  all inequivalent couplings $\tau$ ,
 which is 
the multiple covering of the fundamental domain of the $\tau$ plane.
 Then using the 
fact that the BPS  mass spectrum is finite and the
 superconformal anomaly is a $U(1)$
section , they  uniquely fixed the form of the effective potential. So far their 
procedure is not generalized to higher rank groups.

For $N = 2 $ SUSY Yang-Mills theory with gauge group $G$
 spontaneously broken to 
${U(1)}^r$, where r is the rank of the group, the super conformal Ward identity
implies that \cite{HW},

\bee
\sum_{i =1}^{N_c -1} a_i {\frac{\pa {\cal F}}{\pa a_i}} - 2{\cal F} =
 8~ \pi~ i \bt ~u ,
\ee

where $ {\cal F }$ is the effective action and $ a_i = <{\phi}_i> $,
 $\phi$ is the
scalar 
component of the $N=2$ superfield and $\bt$ is the one loop beta function given as
$ \bt = {\frac{2N_c - N_f}{16 {\pi}^2}}$. Starting from the Seiberg - Witten 
solution
which are described by a family of hyper elliptic curves, and  then using Whitham 
dynamics this relation is also obtained  by Eguchi and Yang \cite{EY}.

For the time being  let us pretend  to forget about the   rich structure of the
SUSY theories due to duality unearthed by Seiberg and Witten. We  only know from 
the conventional quantum field theory the  super
conformal Ward indetity for the $N = 2 $ SUSY Yang Mills theory in its coulomb phase
and only its weak coupling perturbative behaviour. The main objective of our work 
is to derive the exact form of the prepotential starting from this anomaly equation.
We are illustrating it for  the $SU(2)$ gauge group in section 3. and for $SU(3)$
in section 4..First of all we derive a set of second order partial differential
equation in terms of the gauge invariant variables starting from the Ward identity.
The $Z_N$ symmetry, remnant of the $U(1)_R$  symmetry fixes the form of the 
differential equation. The weak coupling result which correspond to the asymptotic
limit is used as the boundary condition which completely fixes the form of the 
differential equation with some unknown coefficients. Then demanding the 
differential equation to be regularly singular all the coefficients are fixed.

    \abschnitt{2. Basic formalism}

We will consider here the  $N = 2$ supersymmetric $SU(N_c)$ gauge theories. For the 
sake 
of simplicity we neglect here flavours. The field content of this theory consists 
of gauge fields $A_{\mu}$, two Weyl fermions $\lambda$ and $\psi$ and a complex 
scalar field $\phi$ all in their adjoint representation of the gauge group $SU(N_c)$.
The theory has $N_c - 1 $ complex dimensional moduli space of vacua which are 
parametrized by the gauge invariant order parameters
\bee
 u_k = {1.\over k} Tr <{\phi}^{k} >, k = 2, .... N_c ,
\ee
 where $\phi$ is the  scalar component of the $N = 2$ chiral super multiplet.
 The moduli space of 
vacua corresponds to the flatness condition $[\phi,\phi^\dagger]=0$ .
 We can always 
rotate $\phi$ in to the caratan subalgebra $\phi = {\sum}_{k = 1} a_{k} H_{k}$ 
with $H_{k} = E_{k,k} - E_{k+1, k+1}$ , and $(E_{k,l})_{ij} = {\delta}_{ik} 
{\delta}_{jl}$. At a generic point the vacuum expectation value of $\phi$ breaks
the $SU(N_c)$ gauge symmetry to ${U(1)}^{N_c - 1}$. The low energy $N =2$   effective 
Lagrangian is written in terms of two chiral  multiplets $( A_i, W_i )$ as

\bee
 {\cal L}_{eff}=Im{\frac{1}{4\pi}}{\bigg\lb}\int d^4\theta\partial_i{\cal F}(A)\bar
 A^i+{1\over2}\int
d^2\theta \partial_i \partial_j {\cal F}(A)W^iW^j ~{\bigg \rb}, 
\ee
where ${\cal F}(A)$ is the holomorphic prepotential and $\partial_i = 
{\frac{\partial}{\partial A_i}}$.
Classically

\bee
{\cal F}_{cl}(A)={\tau\h }\sum_{i=1}^{N_c}{{\bigg\lb} A_i-{1\over N_c}
\sum_{j=1}^{N_c}A_j {\bigg\rb} }^2 
\ee

where 
$\tau = N_c\{{\theta\over2\pi}+{4\pi i\over g^2}\}.$
The one loop perturbative part of the prepotential is given by
\bee
 {\cal F}_1=i\frac{2N_c-N_f}{8\pi}\sum_{i<j}(A_i-A_j)^2
\log{(A_i-A_j)^2\over\Lambda^2}
\ee

Due to $N = 2$ supersymmetry there are no further perturbative correction , however 
there are non-perturbative instaton correction to ${\cal F}$.

 \abschnitt{ 3. $SU(2)$ gauge group}
 
For the sake of clarity of our method we are illustrating the case for $SU(2)$ here.   
However the differential equation approach to this problem is presented by Bilal
\cite{AB} and also pursued by many others.
For $SU(2)$ gauge group the anomalous Ward identity implies 
\bee
a {\frac{\pa F}{\pa a}} - 2 F = 2 ~ i {\frac{u}{\pi}}.
\ee
We define ${\frac{\pa F}{\pa a_i}} = b_i $ which will be interpreted as
the dual of
$a_i$. Taking further derivatives in u we get
\bee
{\frac{{a}^{\pr \pr}}{ a}} =  {\frac{{b}^{\pr \pr}}{ b}} 
   =  - V( u ) 
\ee

which implies
\bea
{a}^{\pr \pr} + V( u ) a & = 0 \nn \\
{b}^{\pr \pr} + V( u ) b & = 0 .&
\eea

Let ${\et}_{1}$ and ${\et}_2$ be two linearly independent  solutions of the 
differential equations. Let ${\cz }= {\frac{{\et}_1}{{\et}_2}}$ be the 
quotient of these two solutions. Very easily it can be shown that
\bea
V( u) =& \h \lb {3\over 2} {(\frac{{\cz}^{\pr \pr}}{{\cz}^{\pr}})}^2 - 
{\frac{{\cz}^{\pr \pr \pr}}{{\cz}^{\pr}}} \rb  \nn \\
 ~ ~ ~ ~  =& \h \{ \cz , u \} &
\eea
Where $\{  \cz , u \}$ is the Schwartzian. 
Let $V(u)$ be analytic in the neighbourhood 
of $u = u_0$ except at $u_0$ itself, then if $\cz $ is continued analytically 
around a closed curve $\cal C$ encircling $u_0$ but no other singularity ,
 we get   
\bea
{\et}_1^{\pr} = &\al {\et}_1 + \bt {\et}_2 \nn \\
{\et}_2^{\pr} = & \gm {\et}_1 + \de {\et}_2 \nn\\
{\cz}^{\pr} = & {\frac{\al \cz + \bt}{\gm \cz + \de }}.&
\eea 
This implies that $\cz$ transforms linearly under this monodromy. 
There is a theorem
which states that if $u( \cz )$ is an automorphic function,
 then in order that the inverse
function $u( \cz )$ be single valued in the neighbourhood of $u_0$ ,
 the transformation 
be an elliptic transformation of angle ${2 \pi}\over p$ 
where $p$ is an integer.
Without going in to details we refer to the reader to Ref. \cite{LF,ZN}, which 
elaborates how to get $V( u )$ .
\bee
V( u ) = {1\over 4} \sum_{i = 1}^n {\bigg\lb} {\frac{1 - {\al}_i^2}{(u -u_i)^2}} +
 { \frac{2 {\bt }_i}{( u - u_i)}} + \gm (u) {\bigg\rb}
\ee
 where ${\al}_i \pi $ is the  internal angle of the $\cz$ polygon 
 at the corner 
homologous with $u_i$ . If $\inft$ is also a singular value of u which 
has an 
angular point of polygon as its homolouge with angle $\ka \pi$ , then
\bee
\sum_i \bt_i =   0  ~ ~,
\sum_i u_i \bt_i =  -\h \sum_i (1 - \al_i^2 ) + \h (1. - \ka^2).
\ee

The classical action posseses $U(1)_R$ symmetry. However the anomaly
and the instatons break this symmetry to $Z_8$ discrete symmetry. Under
this $\phi \ra e^{{i\pi n}\over 2}\phi$, so  for odd n,  ${\phi}^2 \ra  -{\phi}^2$. 
The non-vanishing vacuum expectation value $u = < tr {\phi}^2>$ 
breaks this symmetry further 
to $Z_4$. This also implies that  in the moduli space $u \ra -u $ 
is a symmetry. So
besides $\inft$ we have even number of singularities. 
It is convenient to fix $u_i = u_{i+1}$,
$\al_i = \al_{i+1}$ and $\bt_i = - \bt_{i+1}$. Thus $V( u )$ has
 this specific form
\bee
V( u ) = \h  \sum_{i =1}^n {\bigg{\lb}}    {\frac{(1 - {\al_i}^2)( u^2 + u_i^2) 
}{(u^2 - u_i^2)}} + {\frac{2 u_i \bt_i}{(u^2 - u_i^2)}}{\bigg{\rb}} .
\ee

Around $u \ra \inft$
\bee
V( u ) = {1\over{2 u^2}} \sum_{i =1}^n\lb (1 - {\al_i}^2) + 2 u_i \bt_i \rb
\ee

which is ${\frac{(1 - \ka^2)}{4 u^2}}$ c.f. eq.(12 ). This has the solution
$\et = c~ u^{\h (1 \pm \ka )}$.
 However for $u \ra \inft $ theory is asymptotically free and
the perturbative result is exact which 
gives $ \et_1 = c {\sqrt u}$ 
and $\et_2 = {\sqrt u} \log  u$. This fixes $\ka $ to be 0 .

      To get the solution around an arbitray singular point $u_i$ we make a 
transformation $a = {( u - u_i)}^{\h (1 - \al_i)} f( u )$. This gives rise to
\bee
   f^{\pr \pr} + {\frac{(1 - \al_i)}{( u - u_i)}} f^{\pr}  +
 \h {\frac{\bt_i}{(u - u_i)}} f 
= 0.
\ee

 To solve the above equation  as a power series 
we take $f = {(u - u_i )}^s F( u - u_i)$ , where $F( u - u_i)$ is the power
series  and the 
 indicial equation  is 
\bee
s(s-1) + (1 - \al_i) s = 0
\ee

which gives  $ s= \al_i$. 
 For  $\bt_i$ non-zero   one of the consistent solution will
be for $\al \geq 1$.
 Since maximum value of $\al_i$ is one so we set $\al_i$ to be one.
 So the two solutions are
$ f_1~ = ~ (u - u_i)$ and $f_2~ = ~ {\frac{i}{\pi}}~(u - u_i) \log (u - 
u_i)$.
 Here 
$f_1$ and
$f_2$ can  correspond either to $a$ or $b$ or a
 linear combination of them. We take 
here a most general form of the solution
like $f_1~=~ pb~+~qa$ and $f_2~=~ rb~+sa$ such  that $ps~-~ rq~=~1$.
 They correspond to 
$(p,q)$ and  $(r,s)$ dyons becoming massless  at these singularities.
The monodromy around $u_i$ is given by $(u - u_i) \ra (u - u_i)
e^{2\pi i}$. Under this
\bea
p b^{\pr} + q a^{\pr} =& ~ pb ~ + ~ qa \nn \\
rb^{\pr} + s a^{\pr} =& rb + sa -2 (pb +qa)&
\eea
This gives the monodromy matrix
\bee
M( p, q) = \pmatrix{ 1 + 2pq & 2q^2 \cr -2p^2 & 1 - 2pq}
.\ee  
If $u_i$ is a singularity  also $-u_i$ is  a singularity. The
monodromy around $\inft$
 has to be the composition of all the monodromies in the $u$ plane.
 \bee
 M_{\inft} = M_{-n}\cdots M_{-1}M_1\cdots M_n
 \ee.
 If we chose $M_1$ to be monodromy matrix of a $(1,0)$ dyon which
 is a monopole and $M_{-1}$ to be due to the $(1,1)$ dyon we can very
easily see that
\bee
M_{\inft} = M_{-1}M_{1} = \pmatrix{-1 & 2\cr 0 & -1}
.
\ee
This gives
\bee
M_{\inft} = M_{-n}\cdots M_{-2}M_{\inft}M_{2}\cdots M_n.
\ee

Inorder to satisfy  equation (21),  $(M_2\cdots M_n)^2~ = ~I = (M_{-n}\cdots 
M_{-2})^2~ $ has to be satisfied.
Here  $M_2,\cdots ,M_n$ are all $SL(2,z)$ group elements. Then  
$G_1 = M_2\cdots M_n$ is also an $SL(2,z)$ element. If $G_1^2 = I$ then
either $G_1 = I $ or $G_1 = - I$. Similarly  also
  $G_{-1} =M_{-n}\cdots M_{-2}$  is either $I$ or $ - I$. So the consistency
condition  requires that $G_1 = G_{-1} = I$  or $- I$. Only contractible
loop  has  monodromy which is $I$. This shows that there is no other
singularity besides two strong coupling  singularities  due to (1,0) and
(1,1) dyon and $\inft$.
 However from this  we are not able to algebraically   prove  that 
these three singularities are unique. Since we are  assuming from the 
beginning that  first  two singularites are due to (1,0) and (1,1) dyon.
If we don't assume these singularities
 from the beginning  then we cannot show from the first
principle that  these three are the only singularities.
 There can be some arbitrary finite 
number of monodromy matrices whose product will be $M_{\inft}$. Only
consistency requirement does not give the uniqueness of the three
singularities. Thus  one needs here  more input to show the uniqueness.
This is argued by Magro et al. \cite{Mag}   in a different way
taking  the anomaly to be the $U( 1 )$ section and the finiteness of the
dyon masses.

  \abschnitt{4.  $SU(3)$ case}

   For the $SU(3)$ gauge group  the anomalous Ward indentity gives
\bee
\sum_i  ( \h a_i b_i - {\cal F}) = {3 \over 2}~  i \pi ~ u .
\ee
 
As mentioned earlier the Cartan subalgebra variables are not gauge invariant and we
use $u_k ( a_k)$ as our working variables. We define $ u = u_2$ and $ v = u_3$. For 
$SU(3)$  we have 

\bee
\phi = a_1 H_1 + a_2 H_2
\ee
where 
\bee
H_1 = \pmatrix{ 1 & 0 & 0 \cr 0 & -1 & 0 \cr 0 & 0 & 0 \cr} ~~ and ~~ H_2 =
\pmatrix{ 0 & 0 & 0 \cr 0 & 1 & 0 \cr 0 & 0 & -1 \cr}
\ee
This implies classically $ u = a_1^2 + a_2^2  - a_1 a_2 $ and $ v = a_1 a_2 ( a_1 - 
a_2)$ .

We take dervatives of eq.(22 ) with respect to $u$ and $v$ up to second order
and get a set of second order differential equations as
\bea
\sum_i b_i \pa_{uu}^2 a_i  -  a_i \pa_{uu}^2 b_i = &0 \\
\sum_i b_i \pa_{vv}^2 a_i  -  a_i \pa_{vv}^2 b_i = & 0 \\
\sum_i b_i \pa_{uv}^2 a_i  - a_i  \pa_{uv}^2 b_i = & 0\\
\sum_i b_i \pa_v a_i  - a_i \pa_vb_i =& 0\\
\sum_i \pa_v b_i \pa_u a_i - \pa_u b_i\pa_v a_i = &0 &.
\eea

It is possible to derive differential equations for each $a_i$ and
$b_i$ separately from equations (25)- (29). 
More explicitly taking the sum we can rewrite   e.g. (25) and (26)
as
\bee
a_1 b_1 \lb {1\over b_1} \pa_{uu}^2 b_1 -{1\over a_1}\pa_{uu}^2 a_1\rb =
a_2 b_2 \lb {1\over a_2} \pa_{uu}^2 a_2 - {1\over b_2}\pa_{uu}^2 b_2 \rb ,
\ee
and
\bee
a_1 b_1 \lb {1\over b_1} \pa_{vv}^2 b_1 -{1\over a_1}\pa_{vv}^2 a_1\rb =
a_2 b_2 \lb {1\over a_2} \pa_{vv}^2 a_2 - {1\over b_2}\pa_{vv}^2 b_2 \rb .
\ee

Let us compare the 
ratio  ${\frac{a_1b_1}{a_2 b_2}}$ from equation (30) and (31) to  get

\bee
{\frac{\lb {1\over b_1} \pa_{uu}^2 b_1 -{1\over a_1}\pa_{uu}^2 a_1\rb}
{\lb {1\over b_1} \pa_{vv}^2 b_1 -{1\over a_1}\pa_{vv}^2 a_1\rb}} 
= {\frac{\lb {1\over a_2} \pa_{uu}^2 a_2 - {1\over b_2}\pa_{uu}^2 b_2 \rb }{ \lb 
{1\over a_2} \pa_{vv}^2 a_2 - {1\over b_2}\pa_{vv}^2 b_2 \rb }}
= f_1( u , v ) ,
\ee

where $f_1( u , v)$ is an arbitrary function of $u$ and $v$. Further we
can write
\bee
{1\over b_1}\lb  \pa_{uu}^2 b_1 - f_1 \pa_{vv}^2 b_1\rb = {1\over a_1}\lb
\pa_{uu}^2 a_1 - f_1 \pa_{vv}^2 a_1\rb = f_2(u,v),
\ee
where $f_2(u,v)$ is also an arbitrary function of $u$ and $v$. 
This gives 
\bee
\pa_{uu}^2 a_1 - f_1( u , v)  \pa_{vv}^2 a_1  - f_2(u,v) a_1 = 0.
\ee
and
\bee
\pa_{uu}^2 b_1 - f_1( u ,v ) \pa_{vv}^2 b_1  - f_2(u,v) b_1 = 0.
\ee

Similarly we can get separate equations for  $a_2$ and $b_2$ as
\bee
\pa_{uu}^2 a_2 - f_1(u ,v) \pa_{vv}^2 a_2  - f_3(u,v) a_2 = 0 
\ee
and 
\bee
\pa_{uu}^2 b_2 - f_1(u ,v) \pa_{vv}^2 b_2  - f_3(u,v) b_2 = 0.
\ee
 
Also  from  equations  (25), (27) and (28) we get 
\bee
{\frac{\lb {1\over b_1} \pa_{uu}^2 b_1 -{1\over a_1}\pa_{uu}^2 a_1\rb}
{\lb {1\over a_2} \pa_{uu}^2 a_2 - {1\over b_2}\pa_{uu}^2 b_2 \rb }}
= {\frac{\lb {1\over b_1} \pa_{uv}^2 b_1 -{1\over a_1}\pa_{uv}^2 a_1\rb}
{\lb {1\over a_2} \pa_{uv}^2 a_2 - {1\over b_2}\pa_{uv}^2 b_2 \rb }} 
=  {\frac{\lb {1\over b_1} \pa_{v} b_1 -{1\over a_1}\pa_{v} a_1\rb}
{\lb {1\over a_2} \pa_{v} a_2 - {1\over b_2}\pa_{v} b_2 \rb }}. 
\ee

Now using simple rules of  ratios 
 (i.e. ${\frac{p}{q}}$ = $\frac{r}{s}$  = $\frac{p + t r }{q + t s}$ ) 
and some simple algebraic manipulations we get another set
of separate equations for $a_i$ and $b_i$.
\bea
\pa_{uu}^2 a_1 - g_1( u , v ) \pa_{uv}^2 a_1  - g_2 (u, v)  \pa_v a_1  
 -  g_3(u,v) a_1 = & 0 ,\\ 
\pa_{uu}^2 b_1 - g_1 ( u ,v )\pa_{uv}^2 b_1  - g_2( u, v) \pa_v b_1 - 
 g_3(u,v) b_1 =& 0, \\ 
\pa_{uu}^2 a_2 - g_1(u , v) \pa_{uv}^2 a_2  - g_2 (u ,v) \pa_v a_2 - 
  g_4(u,v) a_2 = & 0 , \\ 
\pa_{uu}^2 b_2 - g_1( u ,v ) \pa_{uv}^2 b_2  - g_2 ( u , v)\pa_v b_2 -
  g_4(u,v) b_2 =& 0. & 
\eea
We note here that there exists various  ways of framing equations
 in various  forms from the Ward identity keeping up to the second derivative
terms.
Here $f_1$, $f_2$ , $f_3$ , $g_1$, $g_2$ ,$g_3$ and  $g_4$ are 
all unknown functions
which will be determined by the argument of symmetry and the existing 
 known solutions in  the semiclassical regime as the boundary condition.
All these equations are selfconsistent and integrable since all these are
derived from the anomalous Ward identity keeping up to the second
derivative terms. So there are no further  constraints on $f_i$ and
$g_i$ for  the equations to  be integrable. These unknowns are
like $V(u)$  of the previous section which in principle can be
algebraically determined.

 We observe that  $a_1$ and $b_1$
obey  the same differential equation and  so also  $a_2$ and $b_2$. In the 
semiclassical limit $u \ra \inft$ and $v \ra \inft$  ( c.f eq.(23 )),
  we have $ a_1^{\pr} = a_1$ and $a_2^{\pr} = a_1 - a_2 $ and 
$a_1^{\pr} = - a_2$ and $a_2^{\pr} = a_1 - a_2$
respectively. This implies that  $a_1$ and $a_2$ obey  the same differential
equation which suggests  $f_2 = f_3$ and $g_3 = g_4$. So we have
\bee
\pa_{uu}^2 {\bf a }- f_1( u , v)~   \pa_{vv}^2 {\bf a}  - f_2(u,v) ~{\bf a } =  0 
\ee
and
\bee
\pa_{uu}^2 {\bf a}  - g_1( u , v )~ \pa_{uv}^2 {\bf a}   - 
g_2 (u, v)~  \pa_v {\bf a}  -   g_3(u,v)~ {\bf a} =  0 
\ee
 
Finally we have a pair of second order 
differential equation for  ${\bf a}$
which has four independent solutions $a_1$, $a_2$, $b_1$ 
and $b_2$. For the case of $SU(2)$ we have only two solutions which ratio gives 
$\cz$ and from there a Schwartzian is emerged which happens to be the potential.
Here also one can form an analogue of $\cz $ which will be 
the   period  matrix,  say $T_{ij}$ in 
terms of
${b_1\over a_1}$, ${b_1\over a_2}$ , ${b_2\over a_1}$ and ${b_2\over a_2}$. 
Any  one  of the solutions can be expressed as a linear combination  of 
 four linearly  independent  solutions; namely  a solution   $\psi = \al_1 \psi_1 +
\al_2\psi_2 + \al_3 \psi_3 + \al_3 \psi_3 $. If $\om$ is a singular point in the
two dimensional $U\times V$ complex space, then a monodromy around $\om$ will change
$\psi_i^{\pr} \ra S_{ij}\psi_j$. $S_{ij}$ can be shown to be $Sp( 4 , Z)$ elements. 
Under this the period matrix  $T_{ij}$  is also transformed under $Sp( 4 , Z)$ 
. An analogue of Schwartzian in higher complex dimensional space can be
found however
the inverse map $\om ( T_{ij} )$ is unknown. This way we cannot proceed to fix
all these unknown function as like as $SU(2)$.
 So we will emphasize on discrete  symmetry
and the  semiclassical boundary condition to recognize  the form of these  unknown 
functions.

 Here $Z_6$ is the remnant of $U(1)_R$ symmetry due to anomaly and instatons on the 
$SU(3)$ moduli space. This implies $\phi \ra e^{\frac{i \pi}{3}}\phi$. In otherwords
$u \ra e^{\frac{ 2 i \pi}{3}}u$ and $v \ra  e^{i \pi}v$. So $u^3$ and $v^2$ are the 
invariant quantities. Of course any constant of mass dimension six is also an
invariant quantity which we denote it as $\Lm^6$ which may coresspond to
the scale parameter of the Wilsonian effective action.
 This dictates the form the differential equation to be
\bee
\pa_{uu}^2 {\bf a } -  f_1(u^3 , v^2,\Lm^6)~ u  ~ \pa_{vv}^2 {\bf a}  
- f_2(u^3 ,v^2 ,\Lm^6)~ u ~{\bf a } =  0 
\ee
and
\bee
\pa_{uu}^2 {\bf a}  - g_1( u^3 , v^2 ,\Lm^6 )~ u^2 v ~ \pa_{uv}^2 {\bf a}  
- g_2 (u^3, v^2, \Lm^6)~ u v~ \pa_v {\bf a}~ - ~  g_3(u^3, v^2 , \Lm^6) ~u 
~{\bf a} =  0. 
 \ee

For $u\ra \inft$ and $v \ra \inft $
we have
\bea
\pa_{uu}^2 {\bf a }\ra & {\frac{\bf a }{4 u^2}}\nn\\
\pa_{vv}^2 {\bf a} \ra & {\frac{\bf a }{4 v^2}}\nn\\
\pa_{uv}^2 {\bf a} \ra & {\frac{\bf a }{4  u v}}\nn\\
\pa_v {\bf a} \ra & {\frac{\bf a }{2  v}}.&
\eea

In the plane when $v = const.$ we shall have
$\pa_{uu}^2 {\bf a } = V ( u ) {\bf a }$ where $V( u )$ is given in
eqn.(13). Similarly this is the case for the plane where $u = const.$
Inorder that all these  conditions be satisfied by equations
((45)-(46)),
we make an ansatz for
\bea
f_1 = & const. = \vp_1 \nn\\
f_2 = & {\frac{\vp_2}{ u^3 + c_1 v^2 + c_2 \Lm^6}}\nn\\
g_1 = & {\frac{\xi_1}{ u^3 + d_1 v^2 + e_1 \Lm^6}}\nn\\
g_2 = & {\frac{\xi_2}{ u^3 + d_2 v^2 + e_2 \Lm^6}}\nn\\
g_3 = & {\frac{\xi_3}{ u^3 + d_3 v^2 + e_3 \Lm^6}} ,&
\eea
where $\vp_1$, $\vp_2$, $\xi_1$, $\xi_2$,
 $\xi_3$, $c_1$, $c_2$, $d_1$, $d_2$,$d_3$
, $e_1$, $e_2$ and $e_3$ are just complex numbers. We note here
that these may  not be  the unique form of these functions. This is merely
our guess since they satisfy the asymptotic conditions in a very simple
way. So many unknown parameters correspond to a large number of
singularities. These numbers are fixed by demanding that the equations
are regularly singular otherwise solutions around an essential singularity
can be  isolated and we cannot recover from that the known results in the
semiclassical regime. The singularity in the semiclassical regime is known
which fixes some of these numbers. We show in detail how this is done in
the sequel.

In the semiclassical limit
$u \ra \inft$ and $v \ra \inft$, we  can neglect $\Lm$.
 Besides separate singularities
due to $u \ra \inft$ and $v \ra \inft$ there is a plane of singularity
when $a_1 = - a_2$, then $u = 3 a^2$ and $v = 2 a^3$ 
and $u^3 = c v^2$ where 
$c = {\frac{27}{4}}$. All $f_i$'s and $g_i$'s are
 singular for $ c_1 = d_1 = d_2 = d_3 = - c$. 
Since this is the only possible singularity in this limit
which  implies that $ c_1 = d_1 = d_2 = d_3 = - c$.

For convenience let us change the variable $u^3 = \al$ and $v^2 = \bt$.
 Now we get
\bee
9 ~\al~ \pa_{\al \al}^2 {\bf a}~ +
~ 6~ \pa_\al {\bf a} - 2~ \vp_1 ( 2~\bt ~\pa_{\bt 
\bt}^2 {\bf a} + ~ \pa_\bt {\bf a}) - {\frac{{\vp}_2}{\al - c \bt}} {\bf a} = 0
\ee
and
\bee
9~ \al ~\pa_{\al \al}^2 {\bf a}~ +~ 6 ~\pa_\al {\bf a} 
- {\frac{6 ~\xi_1 ~\al ~\bt}{\al - c \bt}} \pa_{\al \bt}^2{\bf a} 
 ~ - ~{\frac{2 \xi_2  \bt}{\al - c\bt}}\pa_ \bt{\bf a}  
 ~ - ~{\frac{ \xi_3}{\al - c\bt}}{\bf a} = 0.
\ee

However  these  equations ought to be regularly singular
 for $\al = c\bt$. By 
demanding that we get the following conditions
\bee
9 \al^2 - 4 \vp_1 c^3 \bt^2 - 6 \xi_1 c \al \bt~ = ~ 9 {(\al - c \bt)}^2,
\ee
and
\bee
6 \al + 2 \vp_1 c^2 \bt + 2 \xi_2 c \bt =  6 (\al - c \bt) ,
\ee
and $\vp_2 = 0 $. 
 This gives $\vp_1 = -{\frac{9}{4 c}}$, $\xi_1 = 3 $ and
$\xi_2 = - {\frac{3}{4}}$. When  $u \ra \inft $ and  $ u \geq  v$ 
the asymptotic condition will be satisfied for $\xi_3 = {1\over 4}$ (c.f.
eq.(46 )).  Since  $\vp_2 = 0 $ , from eq. (48) we get $f_2 = 0 $
If we don't take  the equation to be regularly singular then
isolated  singularities can be easily separated from the equation.
Then we can never recover the known  semiclassical result and monodromy
around $\inft$. Thus we get
\bee
\pa_{uu}^2 {\bf a} + {\frac{9}{4 c }} ~ u  ~ \pa_{vv}^2 {\bf a} = 0
\ee 
and
\bee
\pa_{uu}^2 {\bf a} - {\frac{3 u^2 v}{u^3 - c v^2 + e_1 \Lm^6}}\pa_{uv}^2 
{\bf a} + {3\over 4} {\frac{uv}{u^3 - c v^2 + e_2 \Lm^6}}\pa_v {\bf a}
- {\frac{u}{ 4 (u^3 - c v^2 + e_3 \Lm^6)}}{\bf a} = 0.
\ee
Still we have to fix $e_1$ , $e_2$ and $e_3$. For different values of $e_i$'s
we get different  singularities for every term. If really they are 
different then at any such singularity that term can  be separated since all 
 other terms will be  finite. This will correspond to an essential 
singularity for  which in the limit  $\Lm \ra 0 $, then we cannot
recover the 
semiclassical
result. The only legitimate singularities  will correspond to the same 
singularity for all the terms. This implies that $e_1 = e_2 = e_3$. 

To compare our results  with the  
standard Picard-Fuchs equation of Ref. \cite{KLSY}
 we  have to substitute $u = - u$  and take  $e_i = c $ , 
then we will get the exact 
form. Now by changing the variables as $ x = c {\frac{\Lm^6}{u^3}}$ and 
$ y = c{\frac{v^2}{u^3}}$ , equation (53) and  (54) are  written as the 
hypergeometric differential
equation for the Appel system of type $F_4~(a, b, c, d, x, y)$ \cite{AP}.
This equation has been thoroughly  studied in  Ref. \cite{KLSY} in
 three   different regimes of  $u$, $v$ and $\Lm$  which  we don't want
to repeat.
 To compare the results of our
prepotentials with the explicit instaton calculations \cite{Inst} we
take the regime where $u$ is large and $v$ and $\Lm$ are small which
coresspond to  small $x$ and $y$. 
Four independent
solutions are
\bea
a_1&=&\sqrt u F_4\bigg(-\frac{1}{6},\frac{1}{6},1,\frac{1}{2};
\frac{27\Lm^6}{4u^3},\frac{27v^2}{4u^3}\bigg)
+\frac{v}{2u} F_4 \bigg(\frac{1}{3},\frac{2}{3},1,\frac{3}{2};
\frac{27\Lm^6}{4u^3},\frac{27v^2}{4u^3}\bigg)\\
b_1 &=&\frac{1}{2\pi i}
\sqrt u \sum_{m,n}\frac{(-\frac{1}{6})_{m+n}
(\frac{1}{6})_{m+n}}
{m!n!(1)_m(\frac{1}{2})_n}
\bigg(\frac{27\Lm^6}{4u^3}\bigg)^m
\bigg(\frac{27v^2}{4u^3}\bigg)^n \nonumber \\
& &\ \ \times \bigg[
-2\psi(m+1)+\psi(-\frac{1}{6}+m+n)+\psi(\frac{1}{6}+m+n)
+\log\bigg(-\frac{27\Lm^6}{4u^3}\bigg)\bigg] \nonumber \\
& &+\frac{1}{2\pi i}
\frac{v}{2u} \sum_{m,n}\frac{(\frac{1}{3})_{m+n}
(\frac{2}{3})_{m+n}}
{m!n!(1)_m(\frac{3}{2})_n}
\bigg(\frac{27\Lm^6}{4u^3}\bigg)^m
\bigg(\frac{27v^2}{4u^3}\bigg)^n \nn\\
& &\ \ \times \bigg[
-2\psi(m+1)+\psi(\frac{1}{3}+m+n)+\psi(\frac{2}{3}+m+n)
+\log\bigg(-\frac{27\Lm^6}{4u^3}\bigg)\bigg]
\eea
and similarly
\bea
a_2&=&-\sqrt u F_4\bigg(-\frac{1}{6},\frac{1}{6},1,\frac{1}{2};
\frac{27\Lm^6}{4u^3},\frac{27v^2}{4u^3}\bigg)
+\frac{v}{2u} F_4 \bigg(\frac{1}{3},\frac{2}{3},1,\frac{3}{2};
\frac{27\Lm^6}{4u^3},\frac{27v^2}{4u^3}\bigg)
,\\
 b_2 &=&-\frac{1}{2\pi i}
\sqrt u \sum_{m,n}\frac{(-\frac{1}{6})_{m+n}
(\frac{1}{6})_{m+n}}
{m!n!(1)_m(\frac{1}{2})_n}
\bigg(\frac{27\Lm^6}{4u^3}\bigg)^m 
\bigg(\frac{27v^2}{4u^3}\bigg)^n \nonumber \\
& &\ \ \times \bigg[
-2\psi(m+1)+\psi(-\frac{1}{6}+m+n)+\psi(\frac{1}{6}+m+n)
+\log\bigg(-\frac{27\Lm^6}{4u^3}\bigg)\bigg] \nonumber \\
& &+\frac{1}{2\pi i}
\frac{v}{2u} \sum_{m,n}\frac{(\frac{1}{3})_{m+n}
(\frac{2}{3})_{m+n}}
{m!n!(1)_m(\frac{3}{2})_n}
\bigg(\frac{27\Lm^6}{4u^3}\bigg)^m
\bigg(\frac{27v^2}{4u^3}\bigg)^n  \nn\\
& &\ \ \times \bigg[
-2\psi(m+1)+\psi(\frac{1}{3}+m+n)+\psi(\frac{2}{3}+m+n)
+\log\bigg(-\frac{27\Lm^6}{4u^3}\bigg)\bigg]
\eea
where $\psi(x)=\gb^\prime(x)/\gb (x)$
and
\bee
F_4(a,b,c,d;x,y)
=\sum_{m,n}\frac{(a)_{m+n}(b)_{m+n}}{m!n!(c)_m(d)_n}x^my^n.
\ee
where $( a )_m~ = ~ \gb (a + m) / \gb (a)$.
In order to calculate the prepotential ${\cal F}$ we have to  first 
express $u( a_1, a_2)$ and $v ( a_1, a_2)$ as power series in ${\frac{a_1}{\Lm}}$ and
 ${\frac{a_2}{\Lm}}$ by inverting eqn.(54 ) and (56). Then these are substituted in 
$b_1$ and $b_2$. The integration over $a_1$ and $a_2$ gives the prepotential.

We obtained equations (53) and (54) on the basis of $Z_6$ symmetry and
asymptotic conditions.
The most general form of the equations
will be 
\bee
\pa_{uu}^2 {\bf a}  -  
   \sum_i {\bigg\lb}  {\frac{3 u^2 v}{u^3 - c v^2 + e_i \Lm^6}}\pa_{uv}^2
{\bf a} ~  - ~ 
 {3\over 4} {\frac{uv}{u^3 - c v^2 + e_i \Lm^6}}\pa_v {\bf a}
 ~ + ~ {\frac{u}{ 4 (u^3 - c v^2 + e_i \Lm^6)}}{\bf a}{\bigg\rb}   = 0.
\ee
Here again we have multiple strong coupling singularities like the case of $SU(2)$
theory. Again we may  use the argument of monodromy to conjecture that there
are only 
six singularities. However we cannot prove that this is the unique solution.
In this case ordering of monodromy matrices are quite complicated unlike $SU(2)$
case where singularities were paired on the real line. 
\abschnitt{5. Conclusion }
   We are able to derive here an exact prepotential starting from
anomalous
Ward identity of the $N = 2 $ SUSY Yang-Mills theory in
 its broken phase (coulombphase) for  $SU(2)$ gauge group which agrees
with Seiberg-Witten result \cite{SW,Inst}
 and for  $SU(3)$ gauge group which agrees with the results of 
Ref.\cite{KLSY,AF,Inst}.
 We can use this procedure for any 
rank two gauge group. Extensions to higher rank gauge group is possible however it 
will be quite lengthy and tedious. Inclusion of flavours is quite easy for
$SU(2)$ but it will be quite complicated for the  rank two gauge groups.

Although we have not assumed electric-magnetic duality at all however the 
physical interpretation of these singularities will be vanishing condition of
the   $SU( N )$ dyon masses.  We are unable to algebraically  prove here
that the 
Seiberg-Witten solutions \cite{SW,KLSY,AF} are the unique solutions.
 The fact that the monodromy around $\inft$ has to be the 
composition of all the monodromies around all the singularities, in
principle there exists a possibility of finite number of singularities
other than these Seiberg-Witten singularities
 around which the monodromies may  combine to satisfy this condition.
 We fail  to rule out this possibility by imposing only consistency
condition. Somehow it needs another constraint steming from  physical 
condition  to show the uniqueness of the Seiberg-Witten singularities.  

\vfill



\end{document}